\documentclass{elsart}

\usepackage{graphicx}
\usepackage{epsfig}
\usepackage{booktabs}
\usepackage{multirow}

\usepackage{amsmath, amssymb, amsfonts}
\usepackage{color}
\begin{document}

\begin{frontmatter}

% Title, authors and addresses

% use the thanksref command within \title, \author or \address for footnotes;
% use the corauthref command within \author for corresponding author footnotes;
% use the ead command for the email address,
% and the form \ead[url] for the home page:
% \title{Title\thanksref{label1}}
% \thanks[label1]{}
% \author{Name\corauthref{cor1}\thanksref{label2}}
% \ead{email address}
% \ead[url]{home page}
% \thanks[label2]{}
% \corauth[cor1]{}
% \address{Address\thanksref{label3}}
% \thanks[label3]{}

\title{Cosmogenic Activation of Germanium Used for Tonne-Scale Rare Event Search Experiments}
\author[usd]{W.-Z. Wei},
\author[usd,yzu]{D.-M. Mei\corauthref{cor}} and
\corauth[cor]{Corresponding author.}
\ead{Dongming.Mei@usd.edu}
\author[usd]{C. Zhang}

\address[usd]{Department of Physics, The University of South Dakota, Vermillion, South Dakota 57069, USA}
\address[yzu]{School of Physics and Optoelectronic, Yangtze University, Jingzhou 434023, China}

\begin{abstract}
    We report a comprehensive study of cosmogenic activation of germanium used for tonne-scale rare event search experiments. The germanium exposure to cosmic rays on the Earth's surface are simulated with and without a shielding container using Geant4 for a given cosmic muon, neutron, and proton energy spectrum. The production rates of various radioactive isotopes are obtained for different sources separately. We find that fast neutron induced interactions dominate the production rate of cosmogenic activation. Geant4-based simulation results are compared with the calculation of ACTIVIA and the available experimental data. A reasonable agreement between Geant4 simulations and several experimental data sets is presented. We predict that cosmogenic activation of germanium can set limits to the sensitivity of the next generation of tonne-scale experiments. 
\end{abstract}
\end{frontmatter}

\maketitle

\section{Introduction}
The search for rare event physics such as neutrinoless double-beta (0$\nu \beta \beta$) decay~\cite{fta,sde,bil,ag}, i.e. (A, Z)$\rightarrow$(A, Z+2) + 2$\mathrm{e^-}$, and the interactions of dark matter particles~\cite{mar} is of fundamental importance for understanding the physics beyond the Standard Model. Any experimental observation of 0$\nu \beta \beta$ decay would shed light on the intrinsic nature of neutrinos: neutrinos have a Majorana mass component (i.e., neutrinos are their own anti-particles) and the lepton number is violated by two units. The discovery of dark matter particles would help to answer many significant open questions in physics and enable us to have a better understanding of our Universe and its evolution. In order to directly detect such rare physics processes, tonne-scale detectors with high sensitivity and ultra-low background conditions (both internal and external) are required for both 0$\nu \beta \beta$ decay and dark matter experiments.

Due to their excellent energy resolution and high radio-purity, germanium detectors are a well-accepted methodology in direct detection of 0$\nu \beta \beta$ decay and dark matter. The envisioned background level for future germanium-based tonne-scale 0$\nu \beta \beta$ decay experiments is at or below 0.1 count/ton/yr in the 4-keV region of interest (ROI) around the 2039-keV Q-value for $^{76}$Ge 0$\nu \beta \beta$ decay~\cite{hen}. For germanium-based tonne-scale experiments aimed at directly detecting low-mass ($\leq$10 GeV/c$^2$) dark matter particles, the background expectation is about 0.02 events/ton/yr/keV to achieve the desired detector sensitivity~\cite{cdms}. Because of the exposure of germanium detectors to cosmic rays on the Earth's surface, the production of cosmic-ray induced long-lived radioactive isotopes in germanium can be a significant source of background for germanium-based tonne-scale 0$\nu \beta \beta$ decay and dark matter experiments demanding ultra-low background conditions. This necessitates a study evaluating the production rates of cosmogenic isotopes in germanium to have a better quantitative understanding of the effect of cosmic ray induced background on the achievable sensitivity of those two types of experiments. Also, such a study would be of great importance to successfully mitigate the impact of those cosmogenic isotopes.

Compared with ACTIVIA~\cite{activia}, which is a popular C$^{++}$ computer package used in the field of low-background experiments for calculating the cross-section, production rates and decay yields of cosmogenic isotopes based on data tables and semi-empirical formulae, Geant4-based Monte Carlo simulations~\cite{geant_1,geant_2} have been proven to be a more reliable evaluation tool for estimating the cosmogenic production rates in various materials used for xenon-based rare event search experiments~\cite{chao}. This motivates us to estimate the production rates of cosmogenic isotopes in germanium by using Geant4 simulations and ACTIVIA to evaluate which package is more reliable for germanium-based experiments exploring rare event physics by comparing with the available experimental data.

In this paper, the set-up for Geant4 Monte Carlo simulation is presented in Section~\ref{sec:setup}. In Section~\ref{sec:rates}, we show the production rates of several cosmic-ray produced long-lived isotopes in both enriched germanium (for 0$\nu \beta \beta$ decay experiments) and natural germanium (for dark matter experiments) evaluated by Geant4 simulation and ACTIVIA. The comparison between this work and previous estimates and measurements is presented in Section~\ref{sec:comp}. The impact of cosmogenic background on germanium-based tonne-scale rare event search experiments is evaluated in Section~\ref{sec:eva}. Finally, we summarize our conclusions in Section~\ref{sec:conc}.

\section{Geant4 Monte Carlo simulation set-up}
\label{sec:setup}
\begin{figure}
\centering
\includegraphics[angle=0,width=12.cm]{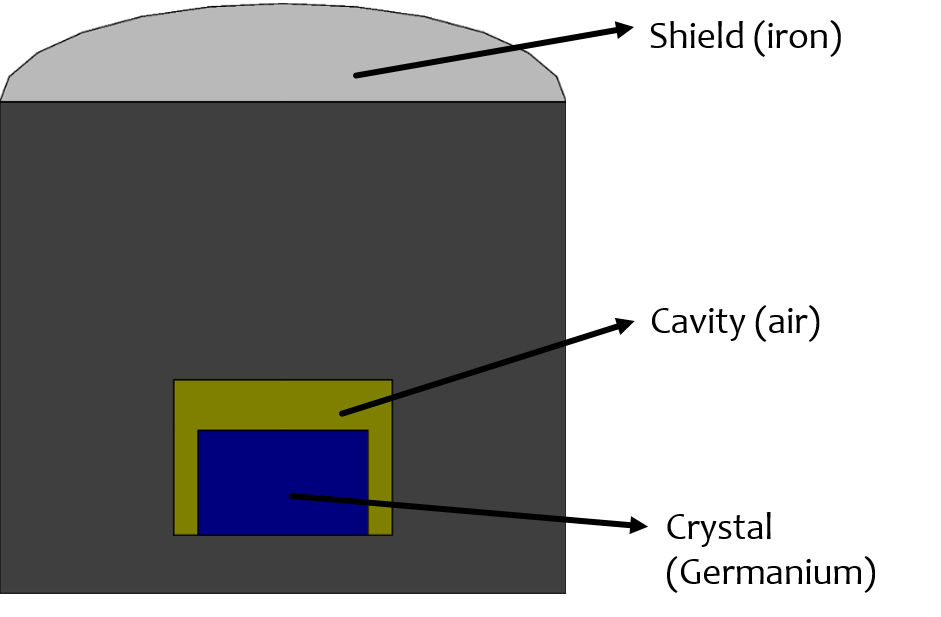}
\caption{\small{Simulated geometry for germanium detector with the iron shield.}}
\label{fig:geo}
\end{figure}

The Geant4 package (V9.5p02 $+$ Shielding physics list) used for this study is the same as the one used in our previous study~\cite{chao}. We simulate two detector geometries in this work. One is a bulk cylinder of germanium with 10 cm in diameter and 10 cm in height. The other is a cylindrical germanium crystal with an iron shield outside as shown in Fig.~\ref{fig:geo}. The dimensions of germanium crystal, air cavity and iron shield shown in Fig.~\ref{fig:geo} are the same as those in the GERDA cosmogenic shield~\cite{bara}, which is shown in Table~\ref{tab:dim}. Note that, within the statistical fluctuations, the size of the crystal is expected to have no impact on the production rates of the cosmogenic isotopes since the simulated production rates will be normalized to a level which is constrained by the total flux of the input energy spectrum and the detector mass. The reason to have two different geometries (with and without the shield) is to quantitatively understand the reduction in the cosmogenic background due to the iron shield. Note that the enriched germanium defined in this simulation is the same as the one used in the MAGE package (a Geant4-based Monte Carlo simulation software framework jointly developed by MAJORANA and GERDA collaborations searching for 0$\nu \beta \beta$ decay of $^{76}$Ge)~\cite{mage}, which is 86.6\% of $^{76}$Ge, 13.1\% of $^{74}$Ge, 0.2\% of $^{73}$Ge and 0.1\% of $^{72}$Ge. 

\begin{table}[tb!!]
\centering
\caption{The dimensions of simulated geometry shown in Fig.~\ref{fig:geo}.}
\label{tab:dim}
\begin{tabular}{|c|c|c|}
\hline \hline
& Diameter (cm) & Height (cm) \\
\hline
 Shield & 140 & 126.5\\
\hline 
 Cavity & 54 & 40\\
\hline
 Crystal & 42 & 27\\
\hline\hline
\end{tabular}
\end{table}

The input energy spectra for this study include the surface muons, slow (energies below 4 MeV) and fast (energies above 4 MeV) neutrons, and protons. The energy spectra of surface muons, slow neutrons and fast neutrons come from the modified Gaisser's formula~\cite{gai,guan}, the air shower simulation results using MCNPX package~\cite{hag2} and the measurements by Gordon~\cite{gordon}, respectively. Our previous work~\cite{chao} has detailed the energy spectrum, energy range, flux and distribution for surface muons and neutrons. For surface protons, the energy spectrum from the work by Hagmann {\it et al.}~\cite{hag1} is used to sample protons right above the simulation geometry. The total flux is normalized to be 0.0001358 cm$^{-2}$s$^{-1}$, which is the total proton flux corresponding to the energy range from 100 MeV to 100 GeV on the surface~\cite{hag1}.

\section{Cosmogenic production rates in germanium at sea level}
\label{sec:rates}
\begin{figure} [htbp]
\includegraphics[angle=0,width=12.cm]{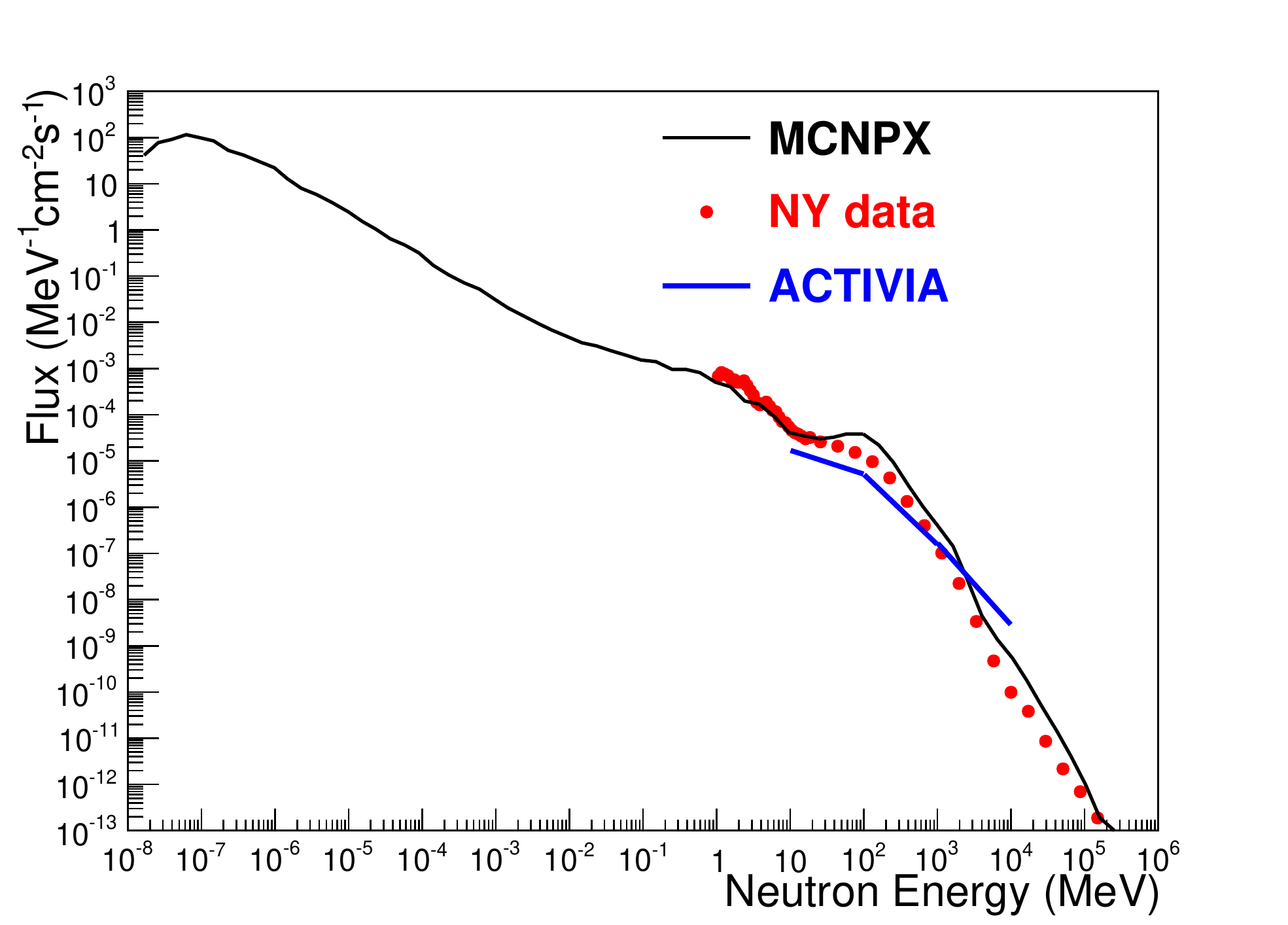}
\caption{The input energy spectra of surface neutrons used in this work. ACTIVIA uses the neutron spectrum at the Earth surface~\cite{arm,geh} (blue curve). The red dots are the measured neutron flux in the New York City (NY data) by Gordon~\cite{gordon}. The black curve is the air shower simulation result from MCNPX simulation code~\cite{hag2}, which is normalized to the NY data. The energy spectra of fast (energies above 4 MeV) and slow (energies below 4 MeV) neutrons used in the Geant4 simulation are from the NY data and MCNPX simulation, respectively.}
\label{fig:neu}
\end{figure}

For both detector geometries (with and without the iron shield), the Geant4-based simulations have been conducted to estimate the production rates of several cosmogenic isotopes in both enriched and natural germanium for a given input energy spectrum of surface muons, neutrons and protons. The simulation results are listed in Table~\ref{tab:enriGe} and Table~\ref{tab:natGe} for enriched and natural germanium, respectively. Also shown in Table~\ref{tab:enriGe} and Table~\ref{tab:natGe} are the production rates calculated by ACTIVIA. Note that the simulated production rates due to slow neutrons are not presented in Table~\ref{tab:enriGe} and Table~\ref{tab:natGe} since slow neutrons have almost no activation in both enriched and natural germanium according to the Geant4 simulation. Also, note that the ACTIVIA package only includes the fast neutron activation of various materials without shielding. Thus, the production rates estimated by ACTIVIA can be compared with Geant4 simulation for fast neutrons only for the case of no shielding. As shown in Fig.~\ref{fig:neu}, the input neutron spectrum~\cite{arm,geh} used by ACTIVIA (blue curve) is different than the one (red dots)~\cite{gordon} used in the Geant4 simulation. Note that cosmic ray neutron flux has a strong dependence on the altitude and the variation at different locations around world has been reported by Gordon~\cite{gordon} and Ziegler~\cite{ziegler}. Ziegler has studied the variation of neutron flux as a function of the altitude. Also, both Ziegler and Gordon found that the variation of the neutron flux from different locations at the sea level caused by geomagnetic rigidity is substantial. In the northern hemisphere, this variation is within 10\%~\cite{gordon}. This work uses the measured cosmic ray neutron flux in the New York City (NY data) from Gordon~\cite{gordon} for fast neutrons, which are the red dots in Fig.~\ref{fig:neu}. To assess the difference between the two neutron spectra used by ACTIVIA and Geant4, it is necessary to conduct the calculation of cosmogenic production rates in germanium using the ACTIVIA package with the default input neutron spectrum (ACTIVIA1) and the one used by Geant4 simulation (ACTIVIA2). 

\begin{table}[tb!!]
\centering
\caption{The production rates (expressed in atoms/kg/day) of cosmogenic isotopes in enriched germanium at sea level estimated by Geant4 simulation (columns 3-10) and ACTIVIA. $n_{fast}$ denotes surface neutrons with energies greater than 4 MeV. $\mu$ and $p$ are surface muons and protons, respectively. Note that the production rates due to slow neutrons are almost zero and therefore, they are not presented in the table. Also, note that the production rates estimated by ACTIVIA can be compared with Geant4 simulation for fast neutrons only for the case of no shielding.}
\label{tab:enriGe}
\begin{tabular}{|p{0.5cm}|p{1cm}|p{0.8cm}|p{0.8cm}|p{0.8cm}|p{0.8cm}|p{0.8cm}|p{0.9cm}|p{0.8cm}|p{0.9cm}|p{2.1cm}|}
\hline\hline
\multicolumn{1}{|c|}{\multirow{2}{1cm}{Isotope}} & \multirow{2}{*}{$t_{1/2}$} &  \multicolumn{2}{c|}{$n_{fast}$} & \multicolumn{2}{c|}{$\mu$} & \multicolumn{2}{c|}{$p$} & \multicolumn{2}{c|}{Total} & \multirow{2}{*}{ACTIVIA1/2} \\ \cline{3-10} 
\multicolumn{1}{|c|}{} &  & no shield & shield & no shield & shield & no shield & shield & no shield & shield &  \\ \hline
 $^{3}_{1}H$& 12.3y  & 47.37 & 8.2 & 0.671 & 0.015 & 5.7 & 0.05 & 53.74 & 8.265 & 33.70/51.27 \\ 
 $^{14}_{6}C$& 5700y  & 0 & 0 & 0 & 0 & 0.003 & 0.0003 & 0.003 & 0.0003 & 0.65/0.21 \\ 
 $^{32}_{14}Si$& 150y & 0.027 & 0 & 0.009 & 0 & 0.009 & 0 & 0.045 & 0 & 0.066/0.027 \\ 
 $^{40}_{19}K$& 1.3e9y & 0.097 & 0.015 & 0 & 0 & 0.04 & 0 & 0.137 & 0.015 & 0.45/0.28 \\ 
 $^{47}_{20}Ca$& 4.5d  & 0.035 & 0 & 0 & 0 & 0.009 & 0 & 0.044 & 0 & 0.016/0.012 \\ 
 $^{46}_{21}Sc$& 83.8d  & 0.19 & 0 & 0.005 & 0 & 0.06 & 0.0009 & 0.255 & 0.0009 & 0.74/0.55 \\ 
 $^{47}_{21}Sc$& 3.3d  & 0.19 & 0.015 & 0.019 & 0 & 0.05 & 0 & 0.256 & 0.015 & 0.33/0.26 \\ 
 $^{44}_{22}Ti$& 63y  & 0.19 & 0.045 & 0 & 0 & 0.06 & 0.0009 & 0.25 & 0.046 & 0.015/0.011 \\ 
 $^{50}_{23}V$& 1.4e17y  & 0.54 & 0.06 & 0.019 & 0 & 0.1 & 0.0006 & 0.66 & 0.06 & 1.46/1.23 \\ 
 $^{51}_{24}Cr$& 27.7d  & 2.13 & 0.27 & 0.042 & 0 & 0.36 & 0.002 & 2.53 & 0.27 & 1.20/1.05 \\ 
 $^{54}_{25}Mn$& 312.3d  & 1.43 & 0.15 & 0.023 & 0.03 & 0.2 & 0.002 & 1.65 & 0.18 & 2.18/2.15 \\ 
 $^{55}_{26}Fe$& 2.7y  & 4.47 & 0.54 & 0.056 & 0 & 0.6 & 0.006 & 5.13 & 0.55 & 1.13/1.18 \\ 
 $^{59}_{26}Fe$& 44.5d  & 1 & 0.1 & 0.014 & 0 & 0.08 & 0.0003 & 1.09 & 0.10 & 0.48/0.59 \\ 
 $^{60}_{26}Fe$& 1.5e6y  & 0.57 & 0.12 & 0 & 0 & 0.04 & 0.001 & 0.61 & 0.12 & 0.23/0.30 \\ 
 $^{56}_{27}Co$& 77.3d  & 0.5 & 0.075 & 0.005 & 0 & 0.07 & 0.0006 & 0.57 & 0.076 & 0.40/0.44 \\ 
 $^{57}_{27}Co$& 271.8d  & 3.32 & 0.45 & 0.074 & 0 & 0.4 & 0.001 & 3.79 & 0.45 & 2.02/2.31 \\ 
 $^{58}_{27}Co$& 70.9d  & 2.91 & 0.49 & 0.028 & 0 & 0.3 & 0.003 & 3.24 & 0.49 & 4.56/5.45 \\ 
 $^{60}_{27}Co$& 5.3y  & 2.35 & 0.34 & 0.014 & 0 & 0.2 & 0.0006 & 2.56 & 0.34 & 3.36/4.41 \\ 
 $^{65}_{30}Zn$& 244.3d  & 24.93 & 5.19 & 0.17 & 0.03 & 1.5 & 0.02 & 26.60 & 5.24 & 5.34/9.66 \\ 
 $^{68}_{32}Ge$& 270d  & 21.75 & 5.26& 0.088 & 0.046 & 1.3 & 0.02 & 23.14 & 5.33 & 7.04/15.38 \\
 \hline\hline
\end{tabular}
\end{table}

\begin{table}[tb!!]
\centering
\caption{The production rates (expressed in atoms/kg/day) of cosmogenic isotopes in natural germanium at sea level estimated by Geant4 simulation (columns 3-10) and ACTIVIA. $n_{fast}$ denotes surface neutrons with energies greater than 4 MeV. $\mu$ and $p$ are surface muons and protons, respectively. Note that the production rates due to slow neutrons are almost zero and therefore, they are not presented in the table. Also, note that the production rates estimated by ACTIVIA can be compared with Geant4 simulation for fast neutrons only for the case of no shielding.}
\label{tab:natGe}
\begin{tabular}{|p{0.5cm}|p{1cm}|p{0.8cm}|p{0.8cm}|p{0.8cm}|p{0.8cm}|p{0.8cm}|p{0.9cm}|p{0.8cm}|p{0.9cm}|p{2.1cm}|}
\hline\hline
\multicolumn{1}{|l|}{\multirow{2}{1cm}{Isotope}} & \multirow{2}{1cm}{$t_{1/2}$} & \multicolumn{2}{c|}{$n_{fast}$} & \multicolumn{2}{c|}{$\mu$} & \multicolumn{2}{c|}{$p$} & \multicolumn{2}{c|}{Total} & \multirow{2}{*}{ACTIVIA1/2} \\ \cline{3-10} 
\multicolumn{1}{|c|}{} &  & no shield & shield & no shield & shield & no shield & shield & no shield & shield &  \\ \hline
 $^{3}_{1}H$&12.3y  & 47.39 & 8.59 & 0.61 & 0.88 & 5.2 & 0.19 & 53.2 & 9.66 & 34.12/52.37 \\ 
 $^{14}_{6}C$& 5700y  & 0 & 0 & 0 & 0 & 0.004 & 0.0009 & 0.004 & 0.0009 & 0.59/0.19 \\ 
 $^{32}_{14}Si$& 150y  & 0.016 & 0 & 0 & 0 & 0.01 & 0.0006 & 0.026 & 0.0006 & 0.053/0.027 \\ 
 $^{40}_{19}K$& 1.3e9y  & 0.086 & 0.015 & 0.014 & 0 & 0.05 & 0.0009 & 0.15 & 0.016 & 0.78/0.54 \\ 
 $^{47}_{20}Ca$& 4.5d  & 0.039 & 0 & 0 & 0 & 0.009 & 0 & 0.048 & 0 & 0.011/0.0087 \\ 
 $^{46}_{21}Sc$& 83.8d  & 0.33 & 0.045 & 0.005 & 0.015 & 0.07 & 0.002 & 0.41 & 0.062 & 0.60/0.49 \\ 
 $^{47}_{21}Sc$& 3.3d  & 0.27 & 0.015 & 0.014 & 0 & 0.06 & 0.001 & 0.34 & 0.016 & 0.24/0.20 \\ 
 $^{44}_{22}Ti$& 63y & 0.49 & 0.03 & 0.009 & 0 & 0.1 & 0.004 & 0.60 & 0.034 & 0.081/0.065 \\ 
 $^{50}_{23}V$& 1.4e17y  & 0.81 & 0.12 & 0.009 & 0 & 0.1 & 0.005 & 0.92 & 0.12 & 1.30/1.21 \\ 
 $^{51}_{24}Cr$& 27.7d  & 3.62 & 0.45 & 0.06 & 0.076 & 0.5 & 0.012 & 4.18 & 0.54 & 2.48/2.50 \\ 
 $^{54}_{25}Mn$& 312.3d & 2.04 & 0.27 & 0.019 & 0.03 & 0.2 & 0.005 & 2.26 & 0.31 & 2.53/2.84 \\ 
 $^{55}_{26}Fe$& 2.7y  & 7.91 & 0.94 & 0.065 & 0.061 & 0.8 & 0.02 & 8.78 & 1.02 & 3.29/4.10 \\ 
 $^{59}_{26}Fe$& 44.5d & 0.9 & 0.19 & 0.014 & 0 & 0.06 & 0.003 & 0.97 & 0.19 & 0.35/0.49 \\ 
 $^{60}_{26}Fe$& 1.5e6y  & 0.49 & 0.09 & 0.005 & 0.03 & 0.02 & 0.001 & 0.51 & 0.12 & 0.16/0.24 \\ 
 $^{56}_{27}Co$& 77.3d & 1.19 & 0.19 & 0.023 & 0.015 & 0.1 & 0.003 & 1.31 & 0.21 & 1.81/2.42 \\ 
 $^{57}_{27}Co$& 271.8d  & 7.38 & 0.94 & 0.074 & 0.061 & 0.7 & 0.026 & 8.15 & 1.03 & 6.38/8.94 \\ 
 $^{58}_{27}Co$& 70.9d  & 5.74 & 0.82 & 0.042 & 0.061 & 0.4 & 0.017 & 6.18 & 0.90 & 7.98/11.37 \\ 
 $^{60}_{27}Co$& 5.3y  & 2.87 & 0.57 & 0.023 & 0.091 & 0.2 & 0.0087 & 3.09 & 0.67 & 2.65/4.06 \\ 
 $^{65}_{30}Zn$& 244.3d  & 75.93 & 16.87 & 0.2 & 0.68 & 2.7 & 0.24 & 78.83 & 17.79 & 19.53/44.19 \\ 
 $^{68}_{32}Ge$& 270d & 182.8 & 46.05 & 1.21 & 3.86 & 4.97 & 0.53 & 188.98 & 50.44 & 10.25/24.65 \\ \hline\hline
\end{tabular}
\end{table}

The following key information is delivered from Tables~\ref{tab:enriGe} and ~\ref{tab:natGe}:
\begin{itemize}
    \item Based on the considerations of total production rates, half life and energy region of interest, $^{60}$Co and $^{68}$Ge produced through the spallation reaction of hadrons with $^{76}$Ge in enriched germanium are the two most prominent cosmogenic radionuclides for $^{76}$Ge 0$\nu \beta \beta$ decay experiments, while $^{3}$H, $^{65}$Zn and $^{68}$Ge in natural germanium are of big concern for germanium-based dark matter detection experiments.
    \item The iron shield can help reduce the cosmogenic isotope production rates by a factor of between 4.34 ($^{68}$Ge) to 283.33 ($^{46}$Sc) for enriched germanium, and a factor of between 3.75 ($^{68}$Ge) to 43.33 ($^{32}$Si) for natural germanium. This demonstrates that an iron shield container is necessary when transporting germanium detectors on the surface.
    \item The cosmogenic activation in both enriched and natural germanium is dominated by fast surface neutrons (energies above 4 MeV). This is because neutron inelastic scattering is the main reaction mechanism of cosmogenic activation in germanium.
    \item  Although the production rates presented in columns 3 and 11 (the second value) were obtained with the same input neutron energy spectrum from the New York data~\cite{gordon}, due to the fact that different cross section libraries are adopted in Geant4 simulation and ACTIVIA, the production rates in enriched germanium estimated by these two packages are different by a factor of 1.08 ($^{3}$H) to 17.27 ($^{44}$Ti), while such difference in natural germanium ranges from a factor of 1.11 ($^{3}$H) to 7.54 ($^{44}$Ti).
    \item The difference in production rates between ACTIVIA1 and ACTIVIA2 caused by a different input neutron spectrum is within a factor of $\sim$3 for both enriched and natural germanium.  
    \item Columns 3 and 11 (the first value) present the production rates estimated by Geant4 and ACTIVIA with its own default input neutron energy spectrum, respectively. The difference in the production rates between these two packages ranges from a factor of 1.25 ($^{56}$Co) to 12.27 ($^{44}$Ti) for enriched germanium, while such difference ranges from a factor of 1.08 ($^{60}$Co) to 17.83 ($^{68}$Ge) for natural germanium. This indicates that, due to different input neutron energy spectrum and cross-section libraries used by Geant4 and ACTIVIA package, the production rates estimated are quite inconsistent with each other for many isotopes. A comparison with the available experimental data is needed to determine which evaluation tool is more reliable for estimating cosmogenic production rates in germanium.
    
\end{itemize} 

\section{Comparison between this work and previous estimates and measurements}
\label{sec:comp}
Tables~\ref{tab:sum1} and ~\ref{tab:sum2} show the comparison between this work and previous production estimates~\cite{bara, edelweiss, avi, miley,klap,back,mei,ceb} and measurements~\cite{edelweiss,avi,elliot,mjd1} for several isotopes in enriched and natural germanium, respectively. As shown in  Tables~\ref{tab:sum1} and ~\ref{tab:sum2}, the production rates of the cosmogenic isotopes considered here vary significantly among different model estimates, ranging from a factor of $\sim$4 ($^{55}$Fe and $^{58}$Co) to $\sim$84 ($^{57}$Co) for enriched germanium, and a factor of $\sim$3 ($^{55}$Fe) to $\sim$27 ($^{57}$Co) for natural germanium. Such large variation in the estimated rates is mainly due to the use of different input cosmic ray neutron spectra and cross-section libraries. Many estimates (~\cite{bara, avi,miley,back}) come from calculations using historical cosmic ray neutron spectra from Ref.~\cite{hess,zig} which are less precise than modern measurements~\cite{gordon,ziegler} used by the estimates in Ref.~\cite{edelweiss,mei,ceb} and this work. 

It is worthwhile mentioning that, as shown in Table~\ref{tab:sum1}, our result for $^{60}$Co in enriched germanium without shielding from Geant4 simulation agree quite well with the available measurement from Ref.~\cite{elliot}, while ACTIVIA1/2 agree with the data within a factor of 2 for $^{60}$Co. For $^{68}$Ge in enriched germanium with shielding, our result from Geant4 simulation agrees within a factor of 2 with the recent preliminary measurement from MAJORANA~\cite{mjd1}. For $^{68}$Ge in enriched germanium without shielding, our results from Geant4 simulation, ACTIVIA1 and ACTIVIA2 are a factor of $\sim$10, $\sim$3.5, and $\sim$7.3 higher than the measurement from Ref.~\cite{elliot}, respectively. Such a difference between our Geant4 simulation and Ref.~\cite{elliot} is likely caused by very high energy neutrons. Since $^{68}$Ge is produced mainly through the spallation reaction of hadrons (neutron part) with $^{76}$Ge, more incident high energy neutrons yield more $^{68}$Ge in enriched germanium. However, the neutrons with different energies used in Ref.~\cite{elliot} were generated from a proton beam, which cannot provide as many high-energy neutrons as the neutron spectrum used in this simulation work. This is why the production rate of $^{68}$Ge estimated in this work from the Geant4 simulation is higher than the measurement in Ref.~\cite{elliot}. 

As shown in Table~\ref{tab:sum2}, our results from the Geant4 simulation agree with the two available measurements~\cite{edelweiss,avi} reasonably well, especially for the three most critical cosmogenic isotopes, $^{3}$H, $^{65}$Zn and $^{68}$Ge, when compared with the most recent measurement by the EDELWEISS detector~\cite{edelweiss}. Note that, for the rate measurement of $^{68}$Ge, EDELWEISS~\cite{edelweiss} only gives a lower limit at a 90\% C.L.. Our results for $^{3}$H from ACTIVIA is also in a reasonable agreement with the EDELWEISS measurement~\cite{edelweiss}. However, for the production of the other two critical isotopes, $^{65}$Zn and $^{68}$Ge, the difference between ACTIVIA and the EDELWEISS data is a factor of more than 2 depending on the neutron flux used in ACTIVIA. For example, for the production of $^{68}$Ge, the difference between ACTIVIA1 and the EDELWEISS data is at least a factor of 7.2 and the difference between ACTIVIA2 and the EDELWEISS data is at least a factor of 3 since ACTIVIA2 used a more accurate neutron spectrum based on the fit to the data~\cite{chao}.
\begin{table}[tb!!]
\centering
\caption{The production rates of several isotopes (expressed in atoms/kg/day) estimated in this work (columns 9 and 10, where G4 and ACT are short for Geant4 and ACTIVIA, respectively) for enriched germanium, compared with previous estimates in Ref.~\cite{bara,avi,miley,back,mei,ceb} and measurements in Ref.~\cite{elliot,mjd1}. The production rates estimated by Geant4 simulation for both geometries (without shielding is (a) and with shielding is (b)) are presented in this comparison. The Geant4 estimates with shielding are to compare with the measurements from Ref.~\cite{mjd1}, and the estimates without shielding are to compare with all previous estimates~\cite{bara, avi,miley,back,mei,ceb} and the measurements~\cite{elliot}.}
\label{tab:sum1}
\begin{tabular}{|p{1cm}|p{0.7cm}|p{0.6cm}|p{0.6cm}|p{0.7cm}|p{0.6cm}|p{0.6cm}|p{0.6cm}|p{0.8cm}|p{1.8cm}|p{1.2cm}|p{1.3cm}|}
\hline\hline
\multirow{3}{*}{Isotope} & \multicolumn{9}{c|}{Calculated Rates} & \multicolumn{2}{c|}{Data} \\ \cline{2-12} 
 & Ref. & Ref. & Ref.& Ref.& Ref. & \multicolumn{2}{c|}{Ref.~\cite{ceb}} & \multicolumn{2}{c|}{This work} & Ref. & Ref. \\ \cline{7-10}
  &~\cite{avi} & ~\cite{miley} & ~\cite{bara}& ~\cite{back}& ~\cite{mei} & ~\cite{ziegler} &~\cite{gordon} & G4 & ACT1/2 & ~\cite{elliot}&~\cite{mjd1}  \\ \hline
\multirow{2}{*}{$^{3}$H}  & \multirow{2}{*}{$\sim$140} & \multirow{2}{*}{} & \multirow{2}{*}{} & \multirow{2}{*}{} & \multirow{2}{*}{24} & \multirow{2}{*}{} & \multirow{2}{*}{} & 47.37(a) & \multirow{2}{*}{33.70/51.27} & \multirow{2}{*}{} & \multirow{2}{*}{140$\pm$10} \\ 
  &  &  &  &  &  &  &  & \multicolumn{1}{l|}{8.265(b)} &  &  &  \\ 
\multirow{2}{*}{$^{54}$Mn}  & \multirow{2}{*}{1.4} & \multirow{2}{*}{} & \multirow{2}{*}{} & \multirow{2}{*}{2.2} & \multirow{2}{*}{0.87} & \multirow{2}{*}{5.4} & \multirow{2}{*}{3.7} & 1.43(a) & \multirow{2}{*}{2.18/2.15} & \multirow{2}{*}{2.0$\pm$1.0} & \multirow{2}{*}{4.4$\pm$4.1} \\ 
  &  &  &  &  &  &  &  & \multicolumn{1}{l|}{0.18(b)} &  &  &  \\ 
\multirow{2}{*}{$^{55}$Fe}  & \multirow{2}{*}{} & \multirow{2}{*}{} & \multirow{2}{*}{} & \multirow{2}{*}{1.4} & \multirow{2}{*}{3.4} & \multirow{2}{*}{2.3} & \multirow{2}{*}{1.6} & 4.47(a) & \multirow{2}{*}{1.13/1.18} & \multirow{2}{*}{} & \multirow{2}{*}{2.1$\pm$0.7} \\ 
  &  &  &  &  &  &  &  & \multicolumn{1}{l|}{0.55(b)} &  &  &  \\ 
$^{57}$Co & 1 & 0.08 &  & 2.9 & 6.7 & 2.3 & 1.7 & 3.32(a) & 2.02/2.31 & 0.7$\pm$0.4 &  \\
$^{58}$Co & 1.8 & 1.6 &  & 5.5 &  & 6.2 & 4.6 & 2.91(a) & 4.56/5.45 &  &  \\ 
$^{60}$Co &  & 3.5 & 3.3 & 2.4 & 1.6 & 6.7 & 5.1 & 2.35(a) & 3.36/4.41 & 2.5$\pm$1.2 &  \\ 
\multirow{2}{*}{$^{65}$Zn} & \multirow{2}{*}{6.4} & \multirow{2}{*}{6} & \multirow{2}{*}{} & \multirow{2}{*}{10.54} & \multirow{2}{*}{20} & \multirow{2}{*}{24} & \multirow{2}{*}{20} & 24.93(a) & \multirow{2}{*}{5.34/9.66} & \multirow{2}{*}{8.9$\pm$2.5} & \multirow{2}{*}{4.3$\pm$3.6} \\ 
  &  &  &  &  &  &  &  & \multicolumn{1}{l|}{5.24(b)} &  &  &  \\ 
\multirow{2}{*}{$^{68}$Ge}  & \multirow{2}{*}{0.94} & \multirow{2}{*}{1.2} & \multirow{2}{*}{5.7} & \multirow{2}{*}{7.6} & \multirow{2}{*}{7.2} & \multirow{2}{*}{13} & \multirow{2}{*}{12} & 21.75(a) & \multirow{2}{*}{7.04/15.38} & \multirow{2}{*}{2.1$\pm$0.4} & \multirow{2}{*}{3.3$\pm$1.6} \\ 
  &  &  &  &  &  &  &  & \multicolumn{1}{l|}{5.33(b)} &  &  &  \\ \hline\hline
\end{tabular}
\end{table}

\begin{table}[tb!!]
\centering
\caption{The production rates of several isotopes (expressed in atoms/kg/day) estimated in this work (columns 12 and 13, where G4 and ACT are short for Geant4 and ACTIVIA, respectively) for natural germanium, compared with previous estimates in Ref.~\cite{bara, edelweiss,avi,miley,klap,back,mei,ceb} and measurements in Ref.~\cite{edelweiss,avi}. Ref.~\cite{edelweiss} estimated the production rates using ACTIVIA based on two different cross sections database. One is from semi-empirical cross sections~\cite{sil1,sil2,sil3,sil4,sil5} (I) and the other is from MENDL-2P libraries~\cite{mend} (II). The production rates estimated by Geant4 simulation in this comparison are based on the geometry without shielding to be consistent with the setup of all other estimates~\cite{bara,edelweiss,avi,miley,klap,back,mei,ceb} and measurements~\cite{edelweiss,avi}.}
\label{tab:sum2}
\begin{tabular}{|p{1cm}|p{0.7cm}|p{0.5cm}|p{0.5cm}|p{0.5cm}|p{0.5cm}|p{0.5cm}|p{0.5cm}|p{0.5cm}|p{0.5cm}|p{0.5cm}|p{0.7cm}|p{1.8cm}|p{1.1cm}|p{1.1cm}|}
\hline\hline
\multirow{3}{*}{Isotope} & \multicolumn{12}{c|}{Calculated Rates} & \multicolumn{2}{c|}{Data} \\ \cline{2-15} 
  & Ref. & Ref. & Ref. & Ref. & Ref. &Ref.& \multicolumn{2}{c|}{Ref.~\cite{ceb}} & \multicolumn{2}{c|}{Ref.~\cite{edelweiss}} & \multicolumn{2}{c|}{This work} & Ref. & Ref.\\ \cline{8-13}
 &~\cite{avi}&~\cite{miley}&~\cite{klap}&~\cite{bara}& ~\cite{back} &~\cite{mei}&~\cite{ziegler}  &~\cite{gordon}&(I)  &(II)  & G4 & ACT1/2 &~\cite{avi}  &~\cite{edelweiss}  \\ \hline
$^{3}$H & $\sim$210 &  &  &  &  & 27.7 &  &  & 46 & 43.5 & 47.39 & 34.12/52.37 &  & 82$\pm$21 \\
$^{54}$Mn & 2.7 &  & 9.1 &  & 2.7 & 2.7 & 7.2 & 5.2 &  &  & 2.04 & 2.53/2.84 & 3.3$\pm$0.8 &  \\ 
$^{55}$Fe &  &  & 8.4 &  & 3.4 & 8.6 & 8 & 6 & 3.5 & 4 & 7.91 & 3.29/4.10 &  & 4.6$\pm$0.7 \\ 
$^{57}$Co& 4.4 & 0.5 & 10.2 &  & 6.7 & 13.5 & 9.7 & 7.6 &  &  & 7.38 & 6.38/8.94 & 2.9$\pm$0.4 &  \\ 
$^{58}$Co & 5.3 & 4.4 & 16.1 &  & 8.5 &  & 13.8 & 10.9 &  &  & 5.74 & 7.98/11.37 & 3.5$\pm$0.9 &  \\
$^{60}$Co&  & 4.8 & 6.6 & 2.9 & 2.8 & 2 & 4.8 & 3.9 &  &  & 2.87 & 2.65/4.06 &  &  \\ 
$^{65}$Zn & 34.4 & 30 & 79 &  & 29 & 37.1 & 77 & 63 & 38.7 & 65.8 & 75.93 & 19.53/44.19 & 38$\pm$6 & 106$\pm$13 \\ 
$^{68}$Ge& 29.6 & 26.5 & 58.4 & 81.6 & 45.8 & 41.3 & 89 & 60 & 23.1 & 45 & 182.8 & 10.25/24.65 & 30$\pm$7 & \textgreater74 \\ \hline\hline
\end{tabular}
\end{table}

\section{Evaluation of the impact of cosmogenic background on future tonne-scale experiments}
\label{sec:eva}
The production rates estimated by Geant4 simulation and ACTIVIA can be converted into radioactivity ($A$), which takes into account the history of exposure and cooling of the detector, through the following formula:
\begin{equation}
    A = R \times (1-e^{-\frac{ln2 \times t_{exp}}{t_{1/2}}})\times e^{-\frac{ln2 \times t_{cool}}{t_{1/2}}},
    \label{eq:radio}
\end{equation}
where, $A$ is the radioactivity, $R$ is the cosmogenic production rate, $t_{1/2}$ is the decay half-live, $t_{exp}$ is the exposure time on the surface and $t_{cool}$ is the cooling time underground.

In this work, we focus on studying the cosmogenic isotopes of interest, as detailed in Section~\ref{sec:rates}. It is worth mentioning that $^{3}$H may have a significant impact on the sensitivity of next generation germanium detectors in direct detection of low-mass dark matter particles considering the following two features of $^{3}$H: 1) The end point energy of $^{3}$H beta decay is only $\sim$18.6 keV. This implies that the entire energy spectrum of $^{3}$H could contribute to the background of low-energy region for low-mass dark matter detection; 2) $^{3}$H has particularly long half-life, $\sim$12.3 years. This indicates that, throughout the entire life of germanium detectors, the activity of $^{3}$H is expected to be essentially the same.

Table~\ref{tab:radioEnriGe} and Table~\ref{tab:radioNatGe} show the radioactivity of those isotopes of interest in enriched and natural germanium, respectively. Note that, in the calculations of the radioactivity shown in Table~\ref{tab:radioEnriGe} and Table~\ref{tab:radioNatGe} by using Eq.~\ref{eq:radio}, the production rate was the total rate from Geant4 simulation in this work (Columns 9 and 10 in Tables~\ref{tab:enriGe} and ~\ref{tab:natGe}), the cooling time ($t_{cool}$) was assumed to be 365 days for all isotopes, and the sea-level cosmic ray exposure time ($t_{exp}$) was assumed to be 30 days for all isotopes with the exception of $^{68}$Ge. For $^{68}$Ge in enriched germanium, considering the optimal case of MAJORANA DEMONSTRATOR~\cite{mjd}, the total effective sea-level exposure is approximately the sum of the following processes: 1) Transportation of the enriched material from Russia to the germanium reduction facility in Oak Ridge, Tennessee, USA ($\sim$40 days, in shielded container); 2) Reduction for $^{76}$GeO$_2$ material at Electrochemical Systems Inc. (ESI) in Oak Ridge and primary zone refinement ($\sim$20 days, no shielded container); 3) Detector fabrication ($\sim$13.5 days, no shielded container); 4) Delivery from Oak Ridge to the Sanford Underground Research Facility (SURF) in lead, South Dakota ($\sim$1.5 days, in shielded container). Thus, $t_{exp}$ = 75 days was assumed in this work to be the exposure time of $^{68}$Ge in enriched germanium. For $^{68}$Ge in natural germanium, the exposure time was assumed to be a factor of three of its decay half-live, {\it i.e.}, $t_{exp}$ = 3$t_{1/2}$. 

\begin{table}[tb!!]
\centering
\caption{The radioactivity (expressed in decays/kg/day) of $^{60}$Co and $^{68}$Ge in enriched germanium calculated through Eq.~\ref{eq:radio} based on the production rates estimated by Geant4 simulation in this work for both geometries, with and without the iron shield. In this calculation, $t_{cool}$ = 365 days for both, $t_{exp}$ = 30 days for $^{60}$Co and $t_{exp}$ = 75 days for $^{68}$Ge.}
\label{tab:radioEnriGe}
\begin{tabular}{|c|c|c|}
\hline \hline
Isotope & no shield & shield \\ \hline 
$^{60}$Co & 1.12$\times$10$^{-2}$ & 1.48$\times$10$^{-3}$  \\ \hline
$^{68}$Ge & 1.58& 0.365  \\ 
\hline \hline
\end{tabular}
\end{table}

\begin{table}[tb!!]
\centering
\caption{The radioactivity (expressed in decays/kg/day) of $^{3}$H, $^{65}$Zn and $^{68}$Ge in natural germanium calculated through Eq.~\ref{eq:radio} based on the production rates estimated by Geant4 simulation in this work for both geometries, with and without the iron shield. In this calculation, $t_{cool}$ = 365 days for all, $t_{exp}$ = 30 days for $^{3}$H and $^{65}$Zn, and $t_{exp}$ = 3($t_{1/2})_{^{68}Ge}$ for $^{68}$Ge.}
\label{tab:radioNatGe}
\begin{tabular}{|c|c|c|}
\hline \hline
Isotope & no shield & shield \\ \hline 
$^{3}$H & 0.232 & 4.21$\times$10$^{-2}$ \\ \hline
$^{65}$Zn & 2.28 & 0.514 \\ \hline
$^{68}$Ge & 64.9 & 17.32 \\ 
\hline \hline
\end{tabular}
\end{table}

\begin{figure} [htbp]
\includegraphics[angle=0,width=12.cm]{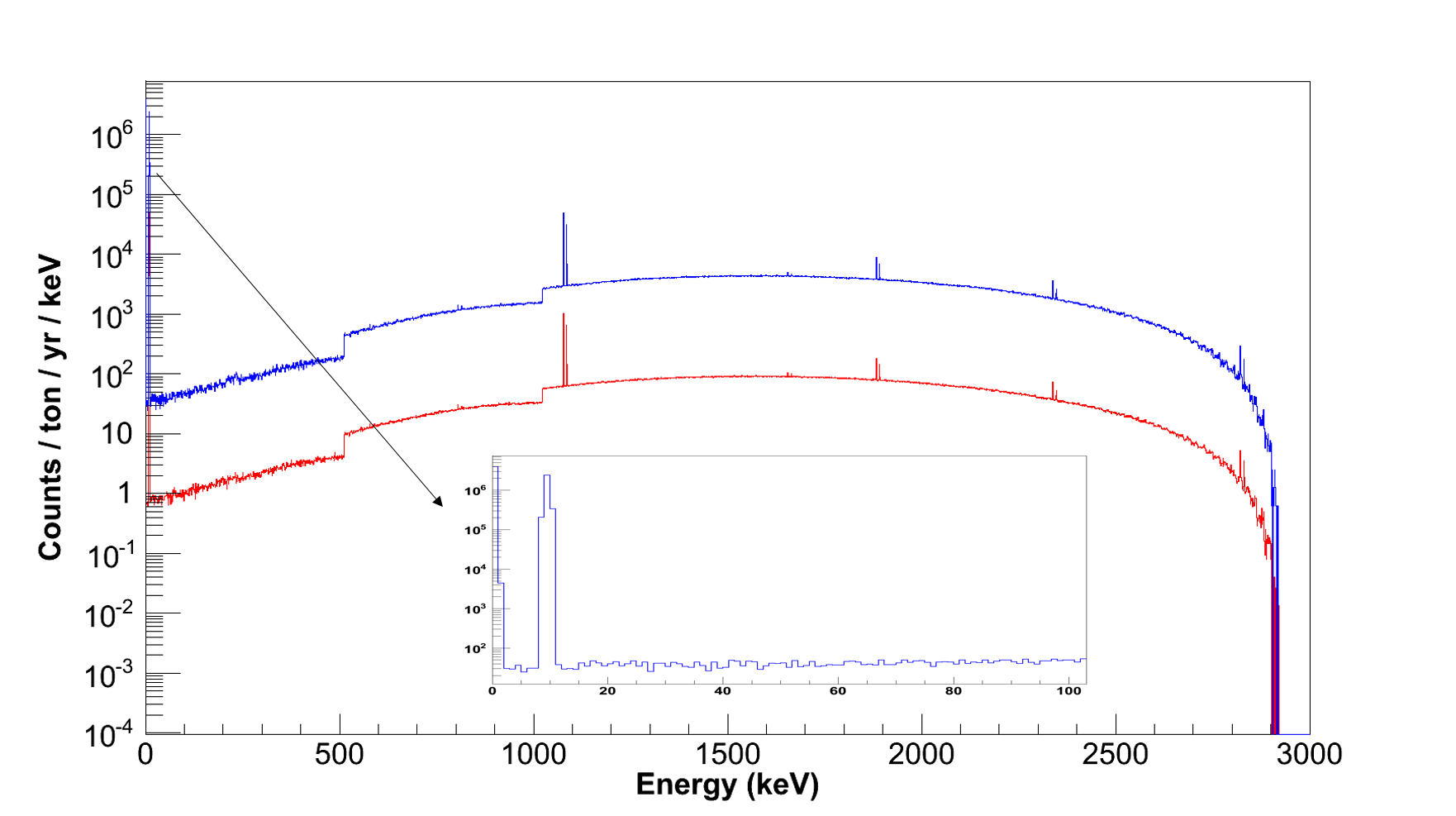}
\caption{The energy spectrum of $^{68}$Ge in enriched (red) and natural (blue) germanium with shielding after normalization. A zoom of the low energy region (0-100 keV) relevant for dark matter searches is also presented for $^{68}$Ge decay in natural germanium.}
\label{fig:ge68_af}
\end{figure}

\begin{figure} [htbp]
\includegraphics[angle=0,width=12.cm]{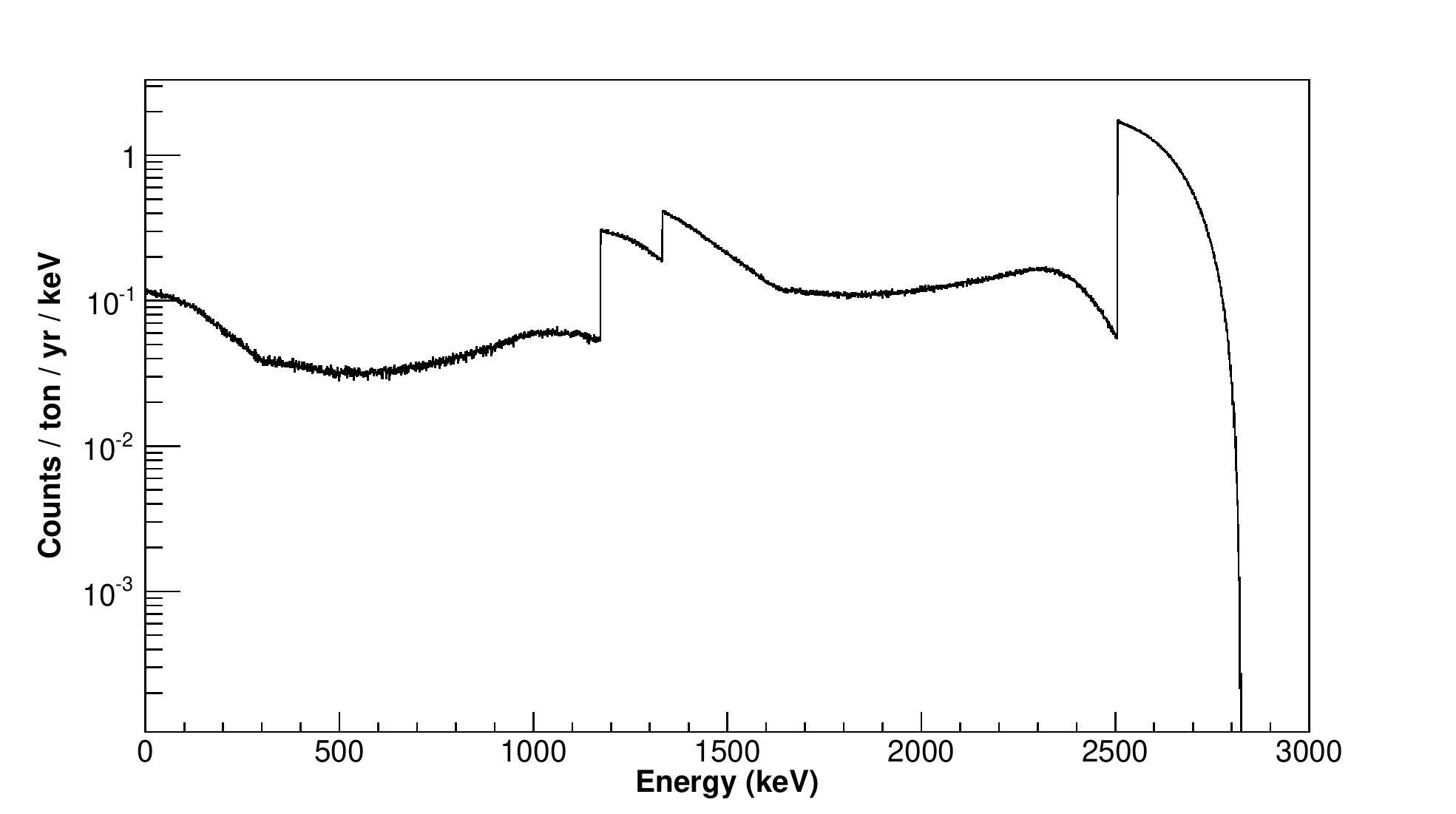}
\caption{The energy spectrum of $^{60}$Co in enriched germanium with shielding after normalization.}
\label{fig:co60_af}
\end{figure}

\begin{figure} [htbp]
\includegraphics[angle=0,width=12.cm]{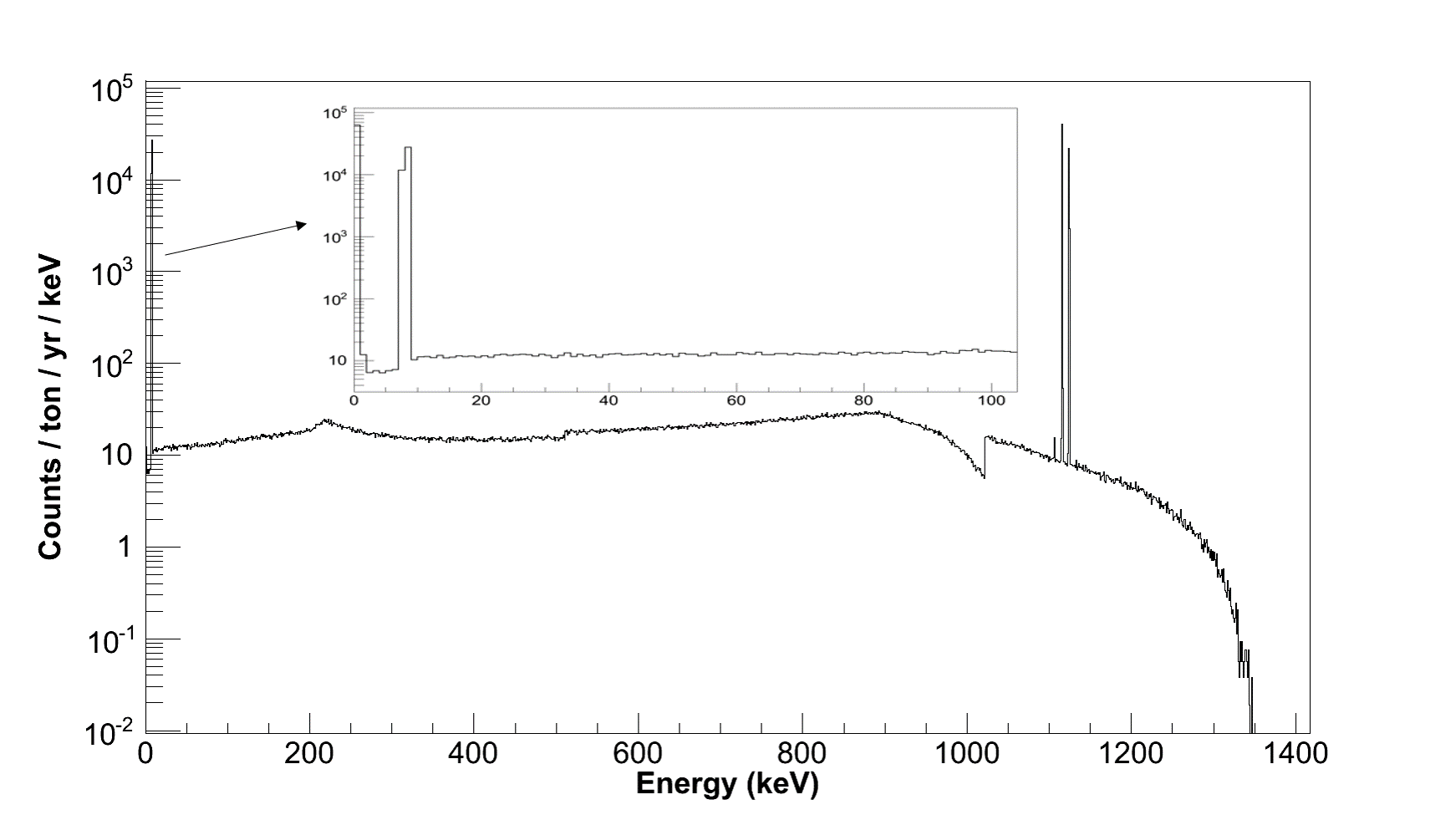}
\caption{The energy spectrum of $^{65}$Zn in natural germanium with shielding after normalization with a zoom of the low energy region (0-100 keV) relevant for dark matter searches.}
\label{fig:zn65_af}
\end{figure}

In order to estimate the contribution, from each cosmogenic isotope shown in Tables~\ref{tab:radioEnriGe} and~\ref{tab:radioNatGe}, to the region of interest of 0$\nu \beta \beta$ decay and dark matter experiments, the energy spectra of those isotopes of interest with geometry shown in Fig.~\ref{fig:geo} were simulated using Geant4 simulation. The Radioactive Decay Mode of Geant4 was used for this simulation. Each isotope was used as the generator in the simulation and assumed to be uniformly distributed inside the germanium detector. The simulated events for each isotope is 10 million. Each simulated energy spectrum is normalized to the radioactivity shown in Tables~\ref{tab:radioEnriGe} and~\ref{tab:radioNatGe} for both cases, with and without the shield. That is, the normalization factor is the ratio of the radioactivity to the total simulated events for each isotope. As examples, we only present the energy spectra of those isotopes of interest in germanium after normalization for the geometry with the shield in this work. That is, the energy spectra of $^{68}$Ge, $^{60}$Co in enriched germanium and $^{68}$Ge, $^{65}$Zn in natural germanium after normalization, which are shown from Fig.~\ref{fig:ge68_af} to Fig.~\ref{fig:zn65_af}. The energy spectra of $^{68}$Ge in enriched and natural germanium are presented in the same figure (Fig.~\ref{fig:ge68_af}) for comparison of any differences between them. The simulation shows that there is no difference between them. Note that the emission of daughter $^{68}$Ga in the decay of $^{68}$Ge was properly included in this simulation. A zoom of the low energy region of interest (0-100 keV) in the spectra of $^{68}$Ge and $^{65}$Zn in natural germanium is presented for the study of dark matter searches. Note that the energy spectrum of $^{3}$H is not shown in this work because the entire energy spectrum of $^{3}$H is in the energy region of interest for low-mass dark matter detection due to the fact that the endpoint energy of $^{3}$H is $\sim$18.6 keV. 

\subsection{The impact of $^{60}$Co and $^{68}$Ge on 0$\nu \beta \beta$ decay experiments}
Table~\ref{tab:res1} presents the simulated counting rates (expressed in counts/ton/yr/ROI) from $^{60}$Co and $^{68}$Ge in the 4-keV ROI around the 2039-keV Q-value for 0$\nu \beta \beta$ decay of $^{76}$Ge before and after cuts. The pulse shape analysis cut (for both $^{60}$Co and $^{68}$Ge) and the time correlation cut (for $^{68}$Ge only) were considered in this work. We assumed that $\sim$90\% ~\cite{mjd} of the events from $^{60}$Co and $^{68}$Ge can be removed by the pulse shape analysis cut, and that $\sim$94.2$\%$ of the events from $^{68}$Ge can be further removed by applying a timing correlation cut. Note that 94.2$\%$ in the time correlation cut is the ratio of the mass of the active region to the total mass of the detector. We assumed that there is a dead layer with thickness 1 mm around a bulk germanium detector with 10 cm in diameter and 10 cm in height. A large number of $^{68}$Ge events can be rejected after a timing correlation cut with the exception of those with the emission of low-energy X-rays deposited in the dead layer. Also note that the detector granularity cut was not considered in this work due to the simple geometry used in this study. Since the granularity depends strongly on the geometry and it is correlated to the pulse shape rejection factor, it is expected that the total rejection factor combining granularity and pulse shape discrimination vary with the number of the detectors in a given geometry.   

\begin{table}[tb!!]
\centering
\caption{The counting rates (expressed in counts/ton/yr/ROI) from $^{60}$Co and $^{68}$Ge in the 4-keV ROI around the 2039-keV Q-value for 0$\nu \beta \beta$ decay experiments before and after cuts.}
\label{tab:res1}
\begin{tabular}{|c|c|c|c|c|c|c|}
\hline \hline
\multicolumn{1}{|c|}{\multirow{2}{*}{Isotope}} & \multicolumn{2}{|c|}{Before cuts} & \multicolumn{2}{|c|}{After pulse shape cut} & \multicolumn{2}{|c|}{After timing correlation cut}\\ \cline{2-7}
\multicolumn{1}{|c|}{} & no shield & shield & no shield & shield & no shield & shield\\ \hline
$^{60}$Co & 3.76 & 0.50 & 0.38 & 0.05 & 0.38 & 0.05\\ \hline
$^{68}$Ge & 1172.6 & 268.9 & 117.3 & 26.9 & 6.8 & 1.6\\ 
\hline \hline
\end{tabular}
\end{table}

As shown in Table~\ref{tab:res1}, the contribution from $^{60}$Co is at a negligible level after a pulse shape cut, while $^{68}$Ge is a concern even after a timing correlation cut for germanium-based 0$\nu \beta \beta$ decay experiments. To understand how much of an effect increasing the enrichment percent of $^{76}$Ge has on the production of $^{68}$Ge, a Geant4 Monte Carlo simulation was carried out with germanium enriched to 100\% in $^{76}$Ge. Based on this simulation, the production rate of $^{68}$Ge is 17.66 kg$^{-1}$day$^{-1}$, which is close to 21.75 kg$^{-1}$day$^{-1}$, the production rate of $^{68}$Ge in germanium enriched to 86.6\% in $^{76}$Ge. Thus, going from 86.6\% to 100\% $^{76}$Ge has little effect on the production rate of $^{68}$Ge. Because of this, there are diminishing returns from enriching past 86.6\% and therefore it is unnecessary.

In order to find out how long a germanium detector needs to be stored underground before taking data to mitigate the impact from $^{68}$Ge, as shown in Table~\ref{tab:res2}, we evaluate the contribution from $^{68}$Ge after a given time period of cooling during which the detector is placed underground. As suggested by Table~\ref{tab:res2}, the cooling time for a germanium detector searching for 0$\nu \beta \beta$ decay can be about 3 years to achieve the background goal of $\sim$0.1 count/ton/yr/ROI, without the granularity and an active veto detector. It is expected that both granularity and an active veto, similar to the GERDA experiment, will further reduce the background events from $^{68}$Ge decay, allowing a much shorter cooling time.

\begin{table}[tb!!]
\centering
\caption{The contribution from $^{68}$Ge in the 4-keV ROI around 2039 keV after a given time period of cooling for the detector, without considering the granularity and an active veto.}
\label{tab:res2}
\begin{tabular}{|c|c|c|}
\hline \hline
\multicolumn{1}{|c|}{\multirow{2}{*}{Cooling time (years)}} & \multicolumn{2}{|c|}{Contribution of $^{68}$Ge (counts/ton/yr/ROI)} \\ \cline{2-3}
\multicolumn{1}{|c|}{} & no shield & shield \\ \hline
1 & 2.7 & 0.6 \\ \hline
2 & 1.1 & 0.2 \\ \hline
3 & 0.4 & 0.1 \\ 
\hline \hline
\end{tabular}
\end{table}

\subsection{The impact of $^{3}$H, $^{65}$Zn and $^{68}$Ge on dark matter experiments}
For germanium-based experiments searching for GeV-scale dark matter particles, the kinetic energy of a recoiling nucleus due to a dark matter interaction is expected to be of orders of a few tens of keV~\cite{meun}. Thus, we focused on the low energy contributions from the three most critical cosmogenic isotopes, $^{3}$H, $^{65}$Zn and $^{68}$Ge for dark matter experiments in this study. Table~\ref{tab:res3} shows the ratios of the number of events in a chosen energy region to that in the entire energy spectra of $^{3}$H, $^{65}$Zn and $^{68}$Ge based on Geant4 simulations. Tables~\ref{tab:res4} and ~\ref{tab:res5} present the simulated number of events (expressed in events/ton/yr) contributed by $^{3}$H, $^{65}$Zn and $^{68}$Ge in those chosen energy regions shown in Table~\ref{tab:res3} for a natural germanium detector with and without the iron shield, respectively. Note that a background rejection factor of 1.0$\times$10$^{-6}$~\cite{cdms} was applied for SuperCDMS-type detectors, while no background rejection power was applied for other germanium detectors used to directly detect dark matter particles. 

\begin{table}[tb!!]
\centering
\caption{The ratio of the number of events in a chosen energy region to that in the entire energy spectra of $^{3}$H, $^{65}$Zn and $^{68}$Ge in natural germanium.}
\label{tab:res3}
\begin{tabular}{|c|c|c|c|}
\hline \hline
Energy region (keV) & $^{3}$H & $^{65}$Zn & $^{68}$Ge \\ \hline
0-5 & 51.51\% & 33.41\% & 31.11\% \\ \hline
0-10 & 86.75\% & 54.24\% & 54.48\% \\ \hline
0-20 & 100\% & 54.31\% & 54.48\% \\ \hline
0-100 & 100\% &54.85\% & 54.51\% \\ 
\hline \hline
\end{tabular}
\end{table}

\begin{table}[tb!!]
\centering
\caption{The contribution (expressed in events/ton/yr) from $^{3}$H, $^{65}$Zn and $^{68}$Ge in a chosen low energy region for natural germanium with the shield. A background rejection factor of 1.0$\times$10$^{-6}$~\cite{cdms} was applied for SuperCDMS-type detectors, while no background rejection power was applied for other types of germanium detectors used in direct detection of dark matter particles.}
\label{tab:res4}
\begin{tabular}{|c|c|c|c|c|c|c|c|c|}
\hline \hline
\multicolumn{1}{|c|}{\multirow{2}{*}{Energy region}} & \multicolumn{4}{|c|}{SuperCDMS-type detectors} & \multicolumn{4}{|c|} {Other germanium detectors} \\ \cline{2-9}
\multicolumn{1}{|c|}{(keV)} & $^{3}$H & $^{65}$Zn & $^{68}$Ge & Total & $^{3}$H & $^{65}$Zn & $^{68}$Ge & Total\\ \hline
0-5 & 0.01 & 0.06 & 3.93 & 4.00 & 7.88$\times$10$^3$ &6.28$\times$10$^4$&3.93$\times$10$^6$ &4.00$\times$10$^6$\\ \hline
0-10 & 0.01 & 0.10 & 6.89 & 7.00 & 1.33$\times$10$^4$ &1.02$\times$10$^5$&6.89$\times$10$^6$ &7.00$\times$10$^6$\\ \hline
0-20 & 0.02 & 0.10 & 6.89 & 7.01 & 1.53$\times$10$^4$& 1.02$\times$10$^5$&6.89$\times$10$^6$ &7.00$\times$10$^6$\\ \hline
0-100 & 0.02 & 0.10 & 6.89 & 7.01 &1.53$\times$10$^4$& 1.03$\times$10$^5$&6.89$\times$10$^6$&7.00$\times$10$^6$\\ 
\hline \hline
\end{tabular}
\end{table}

\begin{table}[tb!!]
\centering
\caption{The contribution (expressed in events/ton/yr) from $^{3}$H, $^{65}$Zn and $^{68}$Ge in a chosen low energy region for natural germanium without shielding. A background rejection factor of 1.0$\times$10$^{-6}$~\cite{cdms} was applied for SuperCDMS-type detectors, while no background rejection power was applied for other types of germanium detectors used in direct detection of dark matter particles.}
\label{tab:res5}
\begin{tabular}{|c|c|c|c|c|c|c|c|c|}
\hline \hline
\multicolumn{1}{|c|}{\multirow{2}{*}{Energy region}} & \multicolumn{4}{|c|}{SuperCDMS-type detectors} & \multicolumn{4}{|c|} {Other germanium detectors} \\ \cline{2-9}
\multicolumn{1}{|c|}{(keV)} & $^{3}$H & $^{65}$Zn & $^{68}$Ge & Total & $^{3}$H & $^{65}$Zn & $^{68}$Ge & Total\\ \hline
0-5 &0.04 & 0.28& 14.70&15.02 &4.36$\times$10$^4$ & 2.78$\times$10$^5$&1.47$\times$10$^7$&1.50$\times$10$^7$ \\ \hline
0-10 &0.07 & 0.45& 25.8&26.32 &7.34$\times$10$^4$ & 4.51$\times$10$^5$&2.58$\times$10$^7$&2.63$\times$10$^7$ \\ \hline
0-20 & 0.08 & 0.45 & 25.8 & 26.33 &8.46$\times$10$^4$&4.52$\times$10$^5$&2.58$\times$10$^7$&2.63$\times$10$^7$\\ \hline
0-100 & 0.08 & 0.46 & 25.8 & 26.34 &8.46$\times$10$^4$&4.56$\times$10$^5$&2.58$\times$10$^7$&2.63$\times$10$^7$ \\ 
\hline \hline
\end{tabular}
\end{table}

As shown in Tables~\ref{tab:res4} and ~\ref{tab:res5}, to reach the background goal of 0.02 events/ton/yr/keV for future germanium-based tonne-scale dark matter experiments, more powerful background rejection techniques are needed even for SuperCDMS-type detectors with the shield. Without shielding, as can be seen from Table~\ref{tab:res5}, $^{65}$Zn and especially $^{68}$Ge are of big concerns. Due to the fact that the half lives of $^{65}$Zn and $^{68}$Ge are not very long (244.3 days and 270 days for $^{65}$Zn and $^{68}$Ge, respectively), one could store germanium detectors for a couple of years before taking data to mitigate the impact of $^{65}$Zn and $^{68}$Ge.

\subsection{Possible solutions to mitigate the impact of cosmogenic isotopes}
There are four possible solutions to reduce the contribution of cosmogenic isotopes in germanium for both 0$\nu \beta \beta$ decay and dark matter experiments to reach their expected detector sensitivities. First, one could increase the amount of iron in the shield container especially on top of the germanium crystal. Previous work by E. Aguayo {\it et al.}~\cite{pnnl} showed that, compared with the current GERDA shield design, the production rate of $^{68}$Ge in enriched germanium can be reduced by a factor of $\sim$30 by increasing the amount of iron to the container’s maximum allowable weight. Secondly, one could process the germanium detectors directly in an underground facility to avoid exposure to cosmogenic activation and the associated hadronic showers above ground~\cite{kla}. If processing germanium detectors has to be above ground, try to limit the exposure time. Take $^{3}$H as an example, according to our calculation, with a fixed value of $t_{cool}$ = 1 yr, the maximum threshold on the exposure time ($t_{exp}$) of germanium crystals with and without shielding to limit the tritium component in order to achieve the background goal (0.02 events/ton/yr/keV) is $\sim$43 days and $\sim$8 days, respectively.  Thirdly, one could use high purity germanium which has already been stored for several years in an underground laboratory~\cite{kla}. Finally, one could build experiments with granularity and active veto to largely reject cosmogenic background events.

\section{Conclusion}
\label{sec:conc}
The cosmogenic activation of germanium for the next generation of rare event search experiments at sea level have been simulated using Geant4 simulations. Muons, fast neutrons, slow neutrons and protons on the Earth's surface are considered individually. The total production rates of several isotopes are compared with ACTIVIA calculations as well as the available experimental data. We find a good agreement between the production rates from Geant4 simulation and experimental data. We conclude that $^{68}$Ge produced in germanium enriched in $^{76}$Ge can be a major source of background for a tonne-scale neutrinoless double-beta decay experiment. Special care, either enriched materials stored in an underground space long before processing into detectors or moving the detectors to an underground site sooner or building experiments with granularity and active veto, must be taken to mitigate the production of $^{68}$Ge in enriched germanium. For a tonne-scale dark matter experiment, $^{3}$H, $^{65}$Zn, and $^{68}$Ge are the main trouble makers. In the case of a SuperCDMS-type detector with excellent n/$\gamma$ discrimination, the sum of these three contributors will limit the sensitivity. For a generic germanium detector without n/$\gamma$ discrimination, cosmogenic production is too high and the only way to reduce it is to grow crystals and make detectors in an underground laboratory. 

\section*{Acknowledgments}
The authors wish to thank Christina Keller and Mitchell Wagner for carefully reading of this manuscript. We are also grateful to Bernhard Schwingenheuer for his comments on an earlier version of the manuscript. This work is supported in part by NSF PHY-0758120, PHYS-0919278, PHYS-1242640, DOE grant DE-FG02-10ER46709, the Office of Research at the University of South Dakota and a governor's research center support by the State of South Dakota. The simulations of this work were performed on High Performance Computing systems at the University of South Dakota.

%
%++++++++++++++++++++++++++++++++++++++++++++++++++++++++++++++++
%------------------------------------------------------------------
%
%\bibliography{apssamp}% Produces the bibliography via BibTeX.

\end{document}